\documentclass{emulateapj}
\usepackage{apjfonts,rotating}

\def\asca{{\itshape ASCA\/}}

\def\chandra{{\itshape Chandra\/}}

\def\nustar{{\itshape NuSTAR\/}}

\def\xray{\hbox{X-ray}}

\def\etal{{et\,al.}}

\def\ltsima{$\; \buildrel < \over \sim \;$}
\def\simlt{\lower.5ex\hbox{\ltsima}}
\def\gtsima{$\; \buildrel > \over \sim \;$}
\def\simgt{\lower.5ex\hbox{\gtsima}}
\def\kms{\ifmmode{~{\rm km~s^{-1}}}\else{~km s$^{-1}$}\fi}
\def\lsim{\lower0.3em\hbox{$\,\buildrel <\over\sim\,$}}
\def\gsim{\lower0.3em\hbox{$\,\buildrel >\over\sim\,$}}

\def\msol{$M_\odot$}
\def\h2{H$_2$}
\def\flux{erg~cm$^{-2}$~s$^{-1}$}
\def\lum{erg~s$^{-1}$}

\def\arcsec{\mbox{$^{\prime\prime}$}}
\def\arcmin{\mbox{$^\prime$}}
\def\sfr{$M_{\odot}$~yr$^{-1}$}

\def\aap{A\&A}
\def\apj{ApJ}
\def\apjl{ApJL}
\def\apjs{ApJS}
\def\aj{AJ}
\def\mnras{MNRAS}

\def\pasj{PASJ}

\begin{document}

\shortauthors{LEHMER ET AL.}
\shorttitle{The Nature of the X-ray Emission in the Nuclear Region of NGC~253}

%
\title{{\it NuSTAR} and {\it Chandra} Insight into the Nature of the 3--40~keV Nuclear Emission in NGC~253}
%

\author{
B.~D.~Lehmer,\altaffilmark{1,2}
D.~R.~Wik,\altaffilmark{2}
A.~E.~Hornschemeier,\altaffilmark{2}
A.~Ptak,\altaffilmark{2}
V.~Antoniou,\altaffilmark{3}
M.~K.~Argo,\altaffilmark{4}
K.~Bechtol,\altaffilmark{5}
S.~Boggs,\altaffilmark{6}
F.~E.~Christensen,\altaffilmark{7}
W.~W.~Craig,\altaffilmark{6,8}
C.~J.~Hailey,\altaffilmark{9}
F.~A.~Harrison,\altaffilmark{10}
R.~Krivonos,\altaffilmark{6}
J.-C.~Leyder,\altaffilmark{2,11}
T.~J.~Maccarone,\altaffilmark{12,13}
D.~Stern,\altaffilmark{14}
T.~Venters,\altaffilmark{2}
A.~Zezas,\altaffilmark{15}
\& 
W.~W.~Zhang\altaffilmark{2}
}

\altaffiltext{1}{The Johns Hopkins University, Homewood Campus, Baltimore, MD 21218, USA}
\altaffiltext{2}{NASA Goddard Space Flight Center, Code 662, Greenbelt, MD 20771, USA} 
\altaffiltext{3}{Department of Physics and Astronomy, Iowa State University, 12 Physics Hall, Ames, IA 50011, USA} 
\altaffiltext{4}{ASTRON, the Netherlands Institute for Radio Astronomy, Postbus 2, 7990 AA, Dwingeloo, The Netherlands}
\altaffiltext{5}{Kavli Institute for Cosmological Physics, Chicago, IL 60637, USA}
\altaffiltext{6}{Space Sciences Laboratory, University of California, Berkeley, CA 94720, USA}
\altaffiltext{7}{DTU Space - National Space Institute, Technical University of Denmark, Elektrovej 327, 2800 Lyngby, Denmark}
\altaffiltext{8}{Lawrence Livermore National Laboratory, Livermore, CA 94720, USA}
\altaffiltext{9}{Columbia Astrophysics Laboratory, Columbia University, New York, NY 10027, USA}
\altaffiltext{10}{Caltech Division of Physics, Mathematics and Astronomy, Pasadena 91125, USA}
\altaffiltext{11}{Universities Space Research Association, Columbia, MD 21044, USA}
\altaffiltext{12}{School of Physics and Astronomy, University of Southampton, Highfield SO17 IBJ, UK}
\altaffiltext{13}{Department of Physics, Texas Tech University, Box 41051, Lubbock, TX 79409-1051, USA}
\altaffiltext{14}{Jet Propulsion Laboratory, California Institute of Technology, Pasadena, CA 91109, USA}
\altaffiltext{15}{Physics Department, University of Crete, Heraklion, Greece}

%
\begin{abstract}
%

We present results from three nearly simultaneous \nustar\ and \chandra\
monitoring observations between 2012 September 2 and 2012 November 16 of the
local star-forming galaxy NGC~253.  The 3--40~keV intensity of 
the inner $\sim$20~arcsec ($\sim$400~pc) nuclear region, as measured by
\nustar, varied by a factor of $\sim$2 across the three monitoring
observations.  The \chandra\ data reveal that the nuclear region contains three
bright \xray\ sources, including a luminous ($L_{\rm 2-10~keV} \sim {\rm few}
\times 10^{39}$ \lum) point source located $\sim$1~arcsec from the dynamical
center of the galaxy (within the 3$\sigma$ positional uncertainty of the
dynamical center); this source drives the overall variability of the nuclear
region at energies $\simgt$3~keV.  We make use of the variability to measure
the spectra of this single hard \xray\ source when it was in bright states.
The spectra are well described by an absorbed ($N_{\rm H} \approx 1.6 \times
10^{23}$~cm$^{-2}$) broken power-law model with spectral slopes and break
energies that are typical of ultraluminous \xray\ sources (ULXs), but not AGN.
A previous \chandra\ observation in 2003 showed a hard \xray\ point source of
similar luminosity to the 2012 source that was also near the dynamical center
($\theta \approx 0.4$~arcsec); however, this source was offset from the 2012
source position by $\approx$1~arcsec.  We show that the probability of the 2003
and 2012 hard X-ray sources being unrelated is $\gg$99.99\% based on the
\chandra\ spatial localizations.  Interestingly, the \chandra\ spectrum of the
2003 source (3--8~keV) is shallower in slope than that of the 2012 hard \xray\
source.  Its proximity to the dynamical center and harder \chandra\ spectrum
indicate that the 2003 source is a better AGN candidate than any of the sources
detected in our 2012 campaign; however, we were unable to rule out a ULX nature
for this source.  Future \nustar\ and \chandra\ monitoring would be well
equipped to break the degeneracy between the AGN and ULX nature of the 2003
source, if again caught in a high state.

%
\end{abstract}
%


%
\section{Introduction}
%

Over the last few decades, imaging and spectroscopy in the \hbox{0.3--10~keV}
bandpass has undergone dramatic improvements thanks to advancements from a
progression of \xray\ observatories.  Due to its proximity and starburst
nature, the nearby galaxy NGC~253 ($D = 3.9$~Mpc based on the tip of the
red-giant branch; Karachentsev \etal\ 2004) has been an ideal target for
studying \xray\ emission from regions more actively star-forming than the Milky
Way and other Local Group galaxies (e.g., Fabbiano \& Trinchieri 1984; Ptak
\etal\ 1997; Vogler \& Pietsch 1999; Pietsch \etal\ 2000, 2001; Strickland
\etal\ 2000; Weaver \etal\ 2002; Bauer \etal\ 2007, 2008).  

Studies of NGC~253 have revealed a diversity of \xray\ emitting populations
throughout the galaxy.  A few dozen \xray\ point sources have been detected
across the optical extent of the disk (e.g., Pietsch \etal\ 2001).  A thin
$\sim$0.4~keV plasma extends several arcminutes along the plane of the disk,
and $\sim$1~keV gas has been observed in a collimated kpc-scale outflow
emanating from the nuclear starburst (e.g., Strickland \etal\ 2000; Bauer
\etal\ 2008; Mitsuishi \etal\ 2012).  The inner nuclear region has been
resolved by \chandra\ into a few bright point sources within a
$\approx$60~arcsec$^2$ region (i.e., the inner $\approx$150~pc).  In this
region, a complex line structure of Fe~K has been resolved into at least three
spectral components due to Fe~{\small I} at 6.4~keV, Fe~{\small XXV} at
6.7~keV, and Fe~{\small XXVI} at 7.0~keV, potentially powered by the
combination of an obscured AGN, supernova (SN) remnants, and/or \xray\ binaries
(Mitsuishi \etal\ 2011).  The point sources include individual \xray\ binaries
and/or the collective emission from sources within star-forming clouds (e.g.,
SN remnants and \xray\ binaries), as well as a hard \xray\ point source
(appearing at energies $\simgt$2~keV) that is located within the 1.2 arcsec
3$\sigma$ uncertainty radius of the dynamical center of the galaxy
(M{\"u}ller-S{\'a}nchez \etal\ 2010).  The central hard \xray\ point-source has
been speculated to be either an obscured ($N_{\rm H} > 10^{23}$~cm$^{-2}$),
low-luminosity ($L_{\rm 2-10~keV} \sim {\rm few} \times 10^{39}$~\lum) AGN or a
super star cluster lit up by SN and/or SN remnants.

High-resolution studies of the nuclear region of NGC~253 at other wavelengths
have revealed a multitude of compact sources within the inner $\sim$200~pc
($\sim$10~arcsec) of the nucleus.  The star-formation rate (SFR) within this
region, based on free-free emission, is estimated to be $\sim$2~\sfr\
(Rodr{\'{\i}}guez-Rico \etal\ 2006), perhaps larger than that of the entire
Milky Way (\hbox{SFR $\approx 1$--2~\sfr}; Hammer \etal\ 2007).   The central
black hole mass has been estimated to be $\approx$$5 \times 10^6$~\msol\ based
on kinematics of the H53$\alpha$ (43~GHz) and H92$\alpha$ (8.31~GHz)
recombination lines within the central 18~pc (0.4~arcsec) of the nucleus
(Rodr{\'{\i}}guez-Rico \etal\ 2006).  Within $\sim$1~arcsec of the dynamical
center are two radio sources, TH2 and TH4, which are separated by 0.36~arcsec
and have been both AGN candidates and candidate markers of the galactic center
(Ulvestad \& Antonucci~1997).  TH2 is the brightest \hbox{15--22~GHz} source in
the region and has historically been adopted as the location of the galactic
center.  The nature of TH2 is still unknown, as the radio spectrum is
consistent with both a low luminosity AGN (LLAGN) and a very compact SN remnant
(e.g., Brunthaler \etal\ 2009).  TH4 has been identified as a water-vapor maser
source (e.g., Hofner \etal\ 2006), which initially suggested an AGN origin.
However, continuum and maser line imaging with the Very Large Array (VLA) and
Very Long Baseline Array (VLBA) showed that there was no strong continuum point
source on arcsecond to milliarcsecond scales, suggesting that the masing source
is most likely to originate from star-formation processes and not a LLAGN
(Brunthaler \etal\ 2009).  Sub-arcsecond near-IR imaging with the Very Large
Telescope (VLT) adaptive optics system found several star-forming regions in
the near vicinity of the nucleus; however, no counterparts were found to TH2
and TH4, again limiting the viability of an AGN presence in NGC~253
(Fern{\'a}ndez-Ontiveros \etal\ 2009).  The above studies have revealed that
the central nuclear region is almost certainly dominated by star formation
processes in the radio and near-IR; however, definitively distinguishing
between the starburst and AGN nature in the \xray\ band has not yet been
possible.

In this paper, we present new insight into the nature of the nuclear region in
NGC~253 as a result of a monitoring campaign consisting of three nearly
simultaneous observations with the {\it Nuclear Spectroscopic Telescope Array}
(\nustar; Harrison \etal\ 2013) and \chandra.  Within the unique energy range
constrained by \nustar, $\approx$10--40~keV for NGC~253, the spectrum
associated with a heavily obscured or low-luminosity AGN will differ
dramatically from non-AGN spectra (e.g., SN, SN remnants, \xray\ binaries, and
ultraluminous \xray\ sources [ULXs]).  Our monitoring campaign therefore
allowed us to simultaneously constrain the spectrum of the central nuclear
region up to $\approx$40~keV and sensitively measure the locations and
contributions of the multiple \xray\ sources associated with the same region
using subarcsecond imaging with \chandra.  The Galactic column density in the
direction of NGC~253 is $1.4 \times 10^{20}$~cm$^{-2}$ (Stark \etal\ 1992).
All \hbox{X-ray} fluxes and luminosities quoted here have been corrected for
Galactic absorption.  At the distance of NGC~253, 1~arcsec corresponds to a
physical distance of 19~pc.  Unless stated otherwise, quoted uncertainties
throughout this paper correspond to 90\% confidence intervals.

%
%
\begin{figure}
\figurenum{1}
\centerline{
\includegraphics[width=8.9cm]{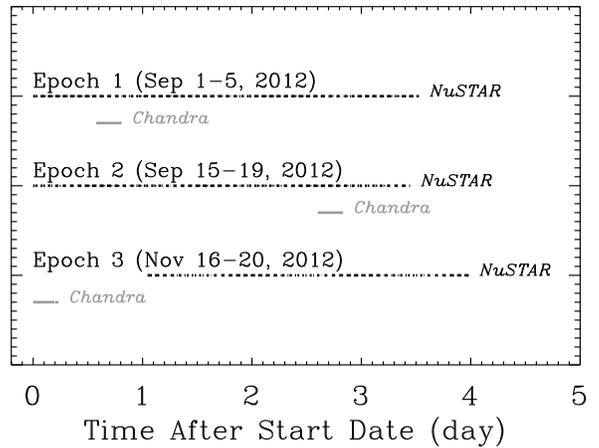}
}
\vspace{0.1in}
\caption{
Relative \nustar\ ({\it black lines\/}) and \chandra\ ({\it gray
lines\/}) observational coverage for each of the three epochs.  For clarity, we
have annotated the total range of observational dates for each epoch.  The
apparently broken up \nustar\ observational intervals are due primarily to
Earth occultations and passages through the SAA.
}
\end{figure}

%
%
\begin{figure*}
\figurenum{2}
\centerline{
\includegraphics[width=17cm]{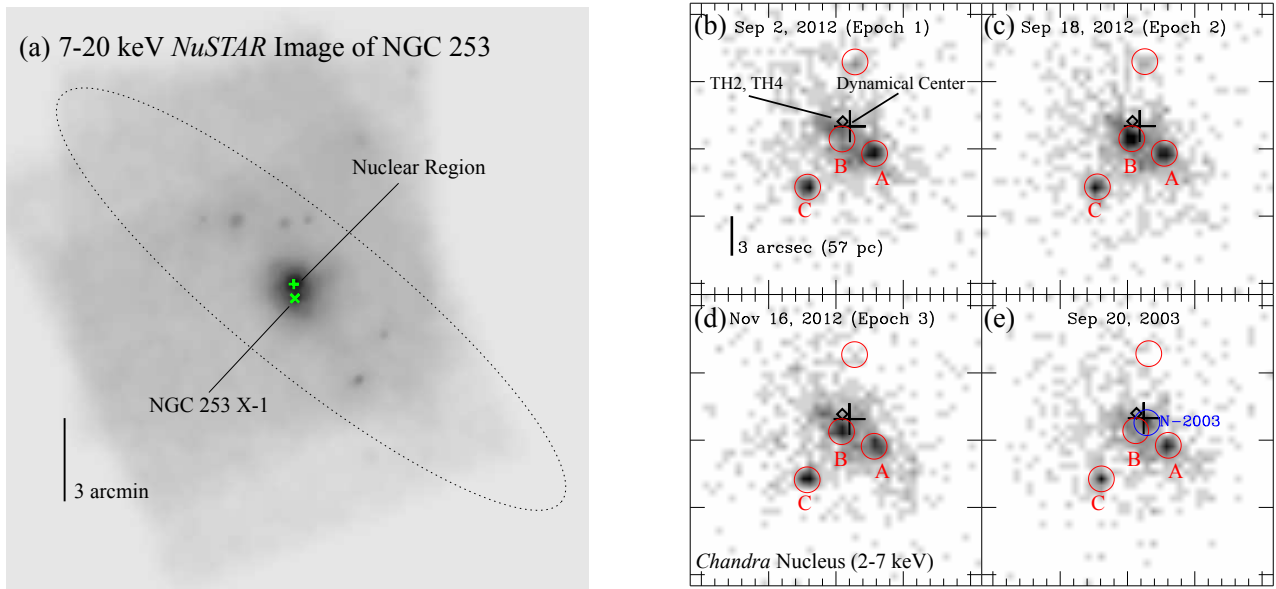}
}
\vspace{0.1in}
\caption{
({\it a\/}) Adaptively-smoothed 7--20~keV \nustar\ image of NGC~253.  The
dotted ellipse represents the optical extent of the galaxy (major axis of
$\approx$24~arcmin) at a surface density of $\mu_B \approx
24$~mag~arcsec$^{-2}$ (Pence~1980).  The locations of the dynamical center
({\it green} ``+'' symbol) and bright off-nuclear source NGC~253 X-1 ({\it
green} ``$\times$'' symbol) have been highlighted.  ({\it b\/})--({\it e\/})
2--7~keV \chandra\ images of the inner $\approx$$22 \times 22$~arcsec$^2$
region of the nucleus for epochs~1, 2, 3, and the 2003 observation (limited to
the first $\approx$20~ks), respectively.  Red circles ($\approx$1~arcsec in
radius) represent the average positions of each \chandra\ source in the three
2012 epochs; the three bright nuclear sources have been labeled ``A''--``C''.
The obscured source found in the nuclear region during our 2012 campaign is source
B, located at ($\alpha$,$\delta$)$_{\rm J2000}$ = 00$^{\rm h}$~47$^{\rm
m}$~33.18$^{\rm s}$, $-25^\circ$~17\arcmin~18.48\arcsec, while the blue circle
in panel~$e$ indicates the location of the obscured source detected in 2003,
``N-2003'' (($\alpha$,$\delta$)$_{\rm J2000}$ = 00$^{\rm h}$~47$^{\rm
m}$~33.12$^{\rm s}$, $-25^\circ$~17\arcmin~17.87\arcsec).  In each panel, the
location of TH2 and TH4 have been shown as a single diamond, and the dynamical
center of the galaxy is indicated by a $2.4$~arcsec~$\times$~2.4~arcsec cross,
corresponding to the 3$\sigma$ astrometric uncertainty (from
M{\"u}ller-S{\'a}nchez \etal\ 2010).
}
\end{figure*}

%
\section{Observations and Data Reduction}
%

Our nearly simultaneous \nustar\ and \chandra\ observations of NGC~253 were
conducted in three observational epochs that began on 2012 September 1, 2012
September 15, and 2012 November 16; hereafter, epochs~1, 2, and 3,
respectively.  During each epoch, we obtained integrated exposures of
$\approx$165~ks with \nustar\ and $\approx$20~ks with \chandra.  In
Figure~1, we show the relative \nustar\ and \chandra\ observational coverage
for the three epochs.  We note that due to Earth occultations and passages
through the South Atlantic Anomaly (SAA), the \nustar\ on target observational
efficiency was 54--63\% (see Harrison \etal\ 2013 for further details).
Therefore, each \nustar\ observation was completed over 260--306~ks (i.e.,
3--3.5~days; see Fig.~1).  For epochs~1 and 2, the shorter \chandra\ exposure
was conducted within the \nustar\ observational interval; however, for epoch~3,
the start of the \chandra\ observation preceded that of the \nustar\
observation by $\approx$1~day.  Additional nearly simultaneous observations
were taken with the VLBA at 1.4~GHz in 8~hr exposures.  However, due to their
depth and frequency, these observations did not yield detections of the
candidate nuclear sources TH2 and TH4, and are therefore not discussed further
here.  Details regarding the VLBA data and NGC~253 galaxy-wide radio population
properties will be presented in a future paper (Argo \etal, in preparation).

Each $\approx$165~ks \nustar\ exposure was conducted using both telescopes, A +
B, which collected \hbox{3--80~keV} photons from the same $12\arcmin \times
12\arcmin$ region centered on the nucleus of NGC~253.  We performed \nustar\
data reduction using {\ttfamily HEASoft 6.12} and {\ttfamily
nustardas}~v.~0.9.0 with {\ttfamily CALDB} v.~20121126.  We processed level~1
data to level~2 products by running {\ttfamily nupipeline}, which performs a
variety of data reduction steps, including filtering out bad pixels, screening
for cosmic rays and observational intervals when the background was too high
(e.g., during passes through the SAA), and projecting accurately the events to
sky coordinates by determining the optical axis position and correcting for the
dynamic relative offset of the optics bench to the focal plane bench due to
motions of the 10~m mast that connects these two benches.  In Figure~2$a$, we
display a \hbox{7--20~keV} \nustar\ image of NGC~253, which was constructed by
merging data from all three observational epochs.

All three of the $\approx$20~ks \chandra\ exposures were conducted using single
\hbox{$16.9\arcmin \times 16.9\arcmin$} ACIS-I pointings (ObsIDs 13830, 13831,
and 13832) with the approximate position of the nucleus set as the aimpoint.
For our data reduction, we used {\ttfamily CIAO}~v.~4.4 with {\ttfamily
CALDB}~v.~4.5.0.  We reprocessed our events lists, bringing level~1 to level~2
using the script {\ttfamily chandra\_repro}, which identifies and removes
events from bad pixels and columns, and filters events lists to include only
good time intervals without significant flares and non-cosmic ray events
corresponding to the standard \asca\ grade set (grades 0, 2, 3, 4, 6).  We
constructed an initial \chandra\ source catalog by searching a
\hbox{0.5--7~keV} image with {\ttfamily wavdetect} (run with a point spread
function [PSF] map created using {\ttfamily mkpsfmap}), which was set at a
false-positive probability threshold of $2 \times 10^{-5}$ and run over seven
scales from 1--8 (spaced out by factors of $\sqrt{2}$ in wavelet scale: 1,
$\sqrt{2}$, 2, 2$\sqrt{2}$, 4, 4$\sqrt{2}$, and 8).  Each initial \chandra\
source catalog was cross-matched to an equivalent catalog, which we created
following the above procedure using a moderately deep ($\approx$80~ks)
\chandra\ ACIS-S exposure from 2003 September 20 (ObsID: 3931).  The 2003
observation is the deepest \chandra\ image available for NGC~253 and has an
aimpoint close to those of the three 2012 observations.  For the purpose of
comparing point sources in the 2012 observations with those of the deep 2003
exposure, we chose to register the 2012 aspect solutions and events lists to
the 2003 frame using {\ttfamily CIAO} tools {\ttfamily reproject\_aspect} and
{\ttfamily reproject\_events}, respectively.  The resulting astrometric
reprojections gave very small astrometric adjustments, including linear
translations of \hbox{$\delta x = -0.49$ to +0.37}~pixels and \hbox{$\delta y =
+$0.28 to 0.37}~pixels, rotations of $-0.026$ to $-0.004$~deg, and pixel scale
stretch factors of 0.999963 to 1.000095.  The final pixel scale of all
observations was 0.492~arcsec per pixel.

Figures~2$b$--2$e$ show 2--7~keV \chandra\ cut-outs of the central nuclear
region of NGC~253 for our 2012 campaign and the 2003 observation.  These images
from our 2012 campaign clearly illustrate that the nuclear region can be
resolved into three bright 2--7~keV point sources, A, B, and C (see
Fig.~2$b$--2$e$), which appear as a single point source with \nustar's
$\approx$18~arcsec full-width half max (FWHM) PSF (see Fig.~2$a$).  The source
located $\approx$1~arcsec to the south of the dynamical center (but within the
3$\sigma$ dynamical center uncertainty radius of 1.2~arcsec), source~B
(($\alpha$,$\delta$)$_{\rm J2000}$ = 00$^{\rm h}$~47$^{\rm m}$~33.18$^{\rm s}$,
$-25^\circ$~17\arcmin~18.48\arcsec), is clearly variable and represents the
best candidate in our 2012 observations for a nuclear point source.  In the
2003 observation (Fig.~2$e$), the nearest nuclear source to source~B (offset by
$\approx$1~arcsec) is labeled ``N-2003'' (($\alpha$,$\delta$)$_{\rm J2000}$ =
00$^{\rm h}$~47$^{\rm m}$~33.12$^{\rm s}$, $-25^\circ$~17\arcmin~17.87\arcsec).
Our analyses below focus on uncovering the nature of source~B and N-2003.

%
%
\begin{figure*}
\figurenum{3}
\centerline{
\includegraphics[width=8.9cm]{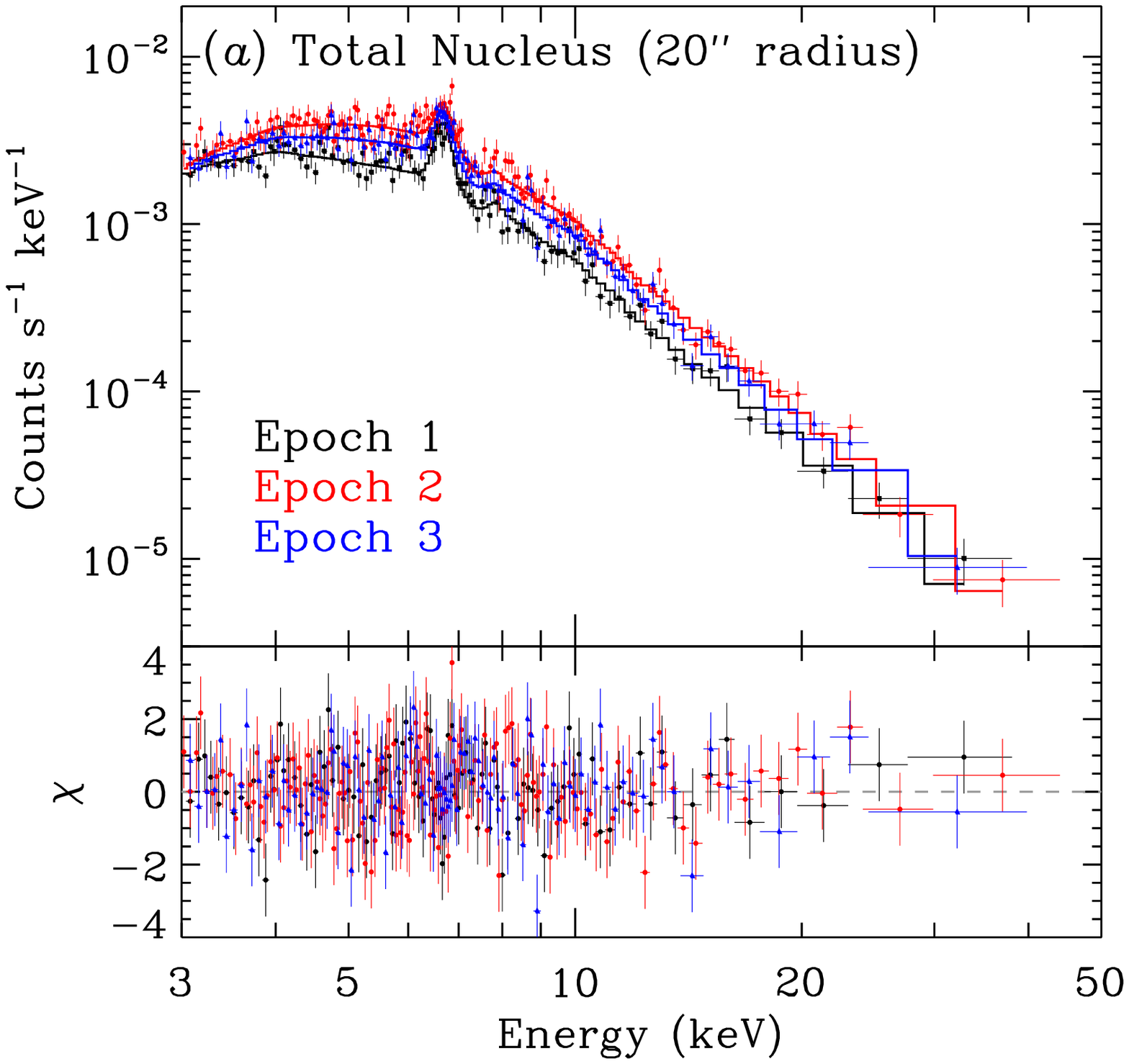}
\hfill
\includegraphics[width=8.9cm]{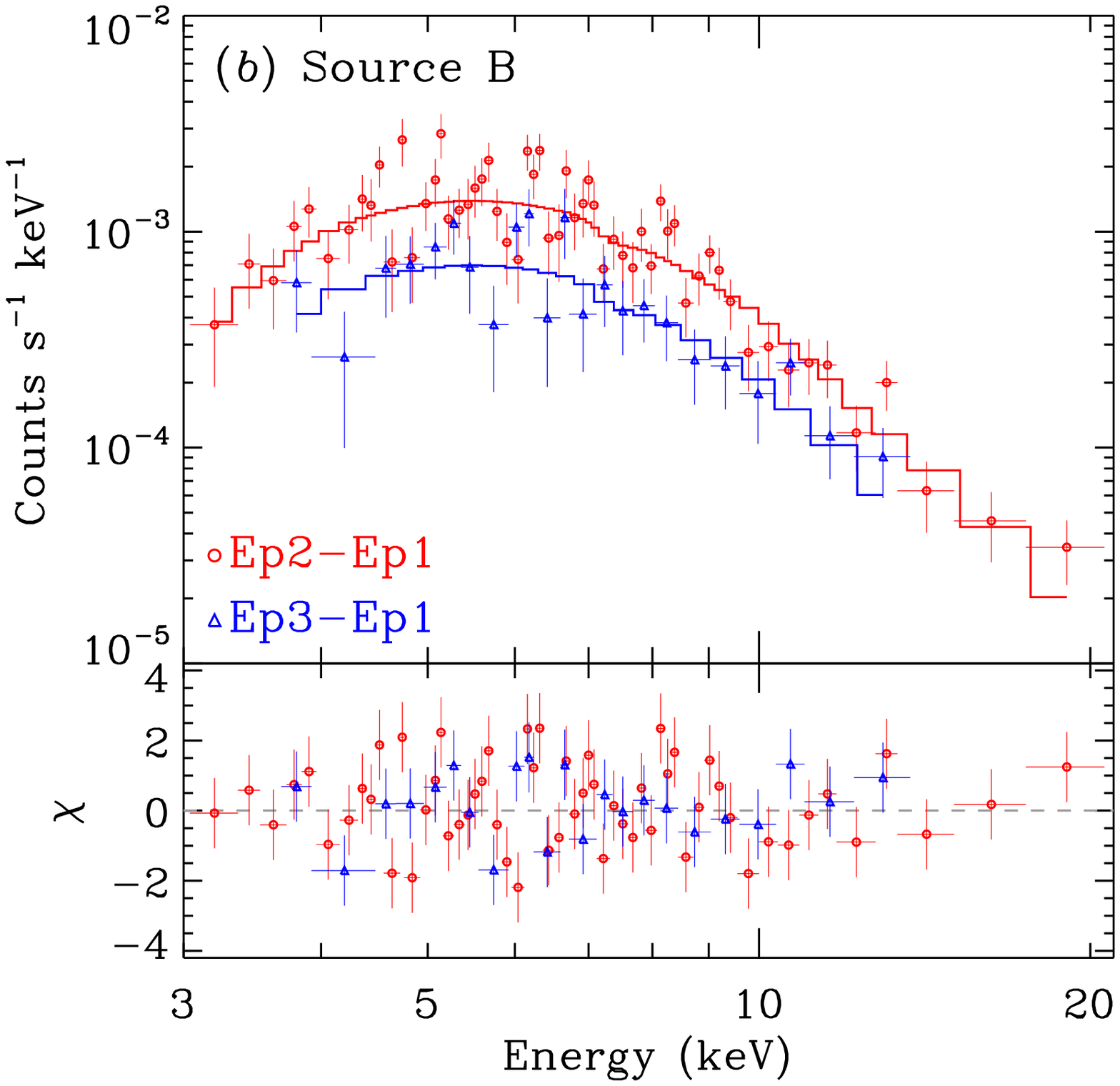}
}
\caption{
({\it a\/})  \nustar\ spectra of the central nuclear region of NGC~253 for the
three observational epochs.  The spectra are all simultaneously fit ({\it
solid curves\/}) by the sum of a non-variable component, including a
$\approx$3.6~keV hot plasma ({\ttfamily apec}) model plus an unabsorbed
{\ttfamily power-law} model ($\Gamma_{\rm non-var} \approx 2.1$), and an
absorbed ($N_{\rm H} \approx 2.5 \times 10^{23}$~cm$^{-2}$) variable {\ttfamily
power-law} component ($\Gamma_{\rm var} \approx 3.0$).  The power-law
components account for emission from persistent and variable \xray\ binaries
and the putative AGN.
({\it b\/}) \nustar\ difference spectra of source~B based on the difference
between epochs~2 and 1 and epochs~3 and 1.  The difference spectra were fit by
a moderately absorbed ($N_{\rm H} \approx 1.6 \times 10^{23}$~cm$^{-2}$) broken
power-law model with $\Gamma_1 \approx 2.4$, $E_{\rm break} \approx 7.9$~keV,
and $\Gamma_2 \approx 3.9$.  Again, the only difference between fits is the
normalization of the power-law component, indicating that the absorption did
not change substantially between observations.  The spectral shape is
consistent with parameters measured for ULXs, but not AGN.
}
\end{figure*}

%
\section{Results and Discussion}
%

%
\subsection{Spectral Properties of the Nuclear Region}

We began by extracting the \nustar\ spectra from the nuclear region of NGC~253
for each of the three observations.  In order to encompass a large fraction of
the source counts in the nuclear region, while minimizing contamination from
nearby unrelated sources, we chose to extract the spectra using circular
regions with radii of 20~arcsec ($\approx$400~pc), somewhat larger than the
\nustar\ half-width at half max (i.e., $\approx$9~arcsec).  This choice of
aperture encompasses 32\% of the \nustar\ PSF (half-power radius 29~arcsec).
Background spectra were extracted from larger apertures (17--37~arcmin$^2$
depending on the observational epoch), which were defined by eye as source-free
regions.  We note that NGC~253 X-1, the bright off-nuclear source
$\approx$30~arcsec to the south of the nucleus, is a ULX with $L_{\rm 2-10~keV}
\approx$~\hbox{(2--3)}~$\times 10^{39}$~\lum\ (based on analysis below) and
provides some non-negligible contribution to the nuclear
region spectra (see Fig.~2$a$).  To estimate these contributions, we extracted
spectra of this source using 20~arcsec radii circular apertures.  The spectra
of NGC~253 X-1 were fit using {\ttfamily xspec}~v.12.7.1 (Arnaud~1996) and a
broken power-law model {\ttfamily bknpower} with varying normalization.  We
found a best-fit low-energy slope of $\Gamma_1 = 2.1 \pm 0.1$, break energy
$E_{\rm break} = 6.8 \pm 0.2$~keV, and high-energy slope $\Gamma_2 = 4.1 \pm
0.1$; no absorption was required in any of these fits.  These values are
similar to those of ULXs studied in detail by Gladstone \etal\ (2009), which
have $L_{\rm 2-10~keV} =$~\hbox{(1--14)}~$\times 10^{39}$~\lum, and are fit
well by broken power-law models with $\Gamma_1 =$~\hbox{1--3}, $E_{\rm break}
=$~\hbox{3--8~keV}, and $\Gamma_2 =$~\hbox{2--7}.  Using a PSF model, we
estimate that the contributions of NGC~253 X-1 to the nuclear region spectra
are small ($\simlt$10\% for all three epochs).   From our best-fit spectral
models for NGC~253 X-1, we constructed fake background spectra corresponding to
its contributions to the nuclear region spectra.  These were used when
performing spectral fits to the nuclear region itself (see below).  To test the
robustness of this procedure, we measured the spectral contributions from the
wings of the {\itshape nuclear region} PSFs to see if they influenced the
extracted spectra of NGC~253 X-1.  Once contributions from the nuclear region
PSFs were subtracted, we found only very small differences (at the $\simlt$7\%
level) in spectral parameters and normalizations derived for NGC~253 X-1.  Such
differences translate into $\simlt$1\% differences in the derived spectral
properties of the nuclear region itself, well below the precision of our
measurements.  We therefore conclude that our procedure for estimating the
contributions of NGC~253 X-1 to the nuclear region is robust.

In Figure~3$a$, we present the \nustar\ spectra of the central nuclear region
for the three epochs.  We do not include the \chandra\ spectra of the nuclear
point sources here, since a direct matching between the multiple components
that lie within the \nustar\ beam is complex and unneccessary for understanding
the hard \xray\ properties of the nuclear region that we are interested in
here.  We know that the \nustar\ emission from the central nuclear region
is a combination of hot plasma, as well as non-variable and variable power-law
sources (i.e., \xray\ binaries and a putative AGN).  We therefore fit the
spectra (with background and NGC~253 X-1 contributions subtracted) of all three
epochs simultaneously using an {\ttfamily apec} plasma model to account for the
hot gas plus two {\ttfamily power-law} components that account for the non-variable
and variable power-law sources, the latter of which we expect to be
primarily due to source~B.  We note that the \xray\ emitting gas is likely to
be complex and contain multiple components (e.g., Pietsch \etal\ 2001;
Mitsuishi \etal\ 2012).  However, since these components have their strongest
influence at energies \hbox{$\simlt$1--2~keV}, we chose to utilize a single hot
plasma to account for the Fe-line emission.  Such a component is expected to
account for hot gas in the core of the starburst associated with SN remnants
(see, e.g., Pietsch \etal\ 2001).  Our adopted model provides a good fit to the
data (see Fig.~3$a$) by simply varying only the normalization of the power-law
associated with variable sources like source~B.  Our best-fit model parameters
include an {\ttfamily apec} plasma with $kT = (3.6 \pm 0.3)$~keV, a non-variable
unobscured power-law component with $\Gamma_{\rm non-var} = 2.1 \pm 0.1$, and
an absorbed ($N_{\rm H} = [2.5 \pm 0.4] \times 10^{23}$~cm$^{-2}$) variable
power-law with $\Gamma_{\rm var} = 3.4 \pm 0.2$.  The 7--20~keV flux of the
nuclear region was factors of $1.69 \pm 0.11$ and $1.39 \pm 0.08$ (1$\sigma$
errors) times higher in epochs~2 and 3, respectively, than its flux in epoch~1
($f_{\rm 7-20~keV}^{\rm epoch~1} = 7.5 \times 10^{-13}$~\flux).

We also tried accounting for the variability by allowing the variable
power-law absorption column, photon index, and normalization components to vary
in different combinations.  Holding the normalization constant and varying the
column density alone provided an acceptable fit and resulted in column
densities $\simgt$$2.9 \times 10^{23}$~cm$^{-2}$ and $\Gamma_{\rm var} \approx
3.7$.  Holding the normalization and varying only the photon index between
epochs did not provide an acceptable fit to the data.  Finally, varying the
column density, photon index, and normalization components simultaneously
provided a good fit to the data, and indicate that the variable component had
high absorption columns ($N_{\rm H, var} \simgt 2.6 \times 10^{23}$~cm$^{-2}$)
and steep photon indices ($\Gamma_{\rm var} \simgt 3.6$).  None of the above
variations improved the quality of the fits to the data over a simple variation
in normalization between epochs.  This analysis indicates that the
\hbox{3--40~keV} spectral slope is much steeper than that found for typical AGN
($\Gamma \approx$~1.5--2.2; e.g., Corral \etal\ 2011; Vasudevan \etal\ 2013).

\subsection{Properties of Source~B}

As discussed in $\S$2 above, the nuclear region of NGC~253 is resolved by
\chandra\ into three bright point sources detected at \hbox{2--7~keV} within a
$\approx$60~arcsec$^2$ region.  The \nustar\ PSF is too large to distinguish
spatially these sources.  However, the \nustar\ and \chandra\ spectral
coverages overlap in the energy range of $\approx$3--7~keV, allowing us to
measure indirectly the levels that each source contributes to the \nustar\ flux
variations.  With \chandra, we found that only the central source (labeled
source ``B'' in Fig.~2$b$--2$e$) varied significantly over the three epochs.
To estimate the fractional contributions that source~B provided to the total
\nustar\ extraction region spectra, we measured 4--7~keV count-rates from
source~B and a larger nuclear region using circular apertures of 1.0~arcsec and
15~arcsec, respectively.  We find that the 4--7~keV flux from the larger
nuclear region was factors of $1.6 \pm 0.1$ and $1.2 \pm 0.1$ (1$\sigma$
errors) higher in epochs~2 and 3, respectively, compared with epoch~1; this is
consistent with the \nustar\ observations over the same energy range, which
were $1.58 \pm 0.07$ and $1.33 \pm 0.06$ (1$\sigma$ errors), respectively.
Given that the shorter 20~ks (0.23~day) \chandra\ exposures did not cover the entire $\approx$3--3.5~day periods spanned by the \nustar\ observations (see Fig.~1), this result
indicates that the nuclear region was unlikely to vary substantially during
each of the observational epochs.  We verified this result by inspecting \nustar\
light curves of the nuclear region.  These light curves were constructed by
extracting events lists from a 10~arcsec radius circle centered on the nucleus
and measuring the mean count-rates of this region in 20~ks bins.  We found no
signatures of strong variability within each of the three observational epochs.
A K-S test revealed that each of the nuclear region light curves were
consistent with those expected from constant-intensity sources.  We therefore
conclude that the nuclear region did not vary substantially on timescales of
$\approx$6--48~hr.  Our \chandra\ observations indicate that source~B itself
was brighter in epochs~2 and 3 compared with epoch~1 by factors of $5.2 \pm
1.0$ and $2.6 \pm 0.6$ (1$\sigma$ errors), respectively.  Once we subtract the
contributions of source~B to the total \chandra\ fluxes, contained within the
larger \nustar\ extraction region, we find no significant remaining variations
in the 4--7~keV non-source~B emission between all three epochs, indicating that
nuclear region variability observed within the \nustar\ PSF, at least in the
4--7~keV band, are likely entirely attributable to variations in source~B.  

%
%
\begin{figure*}
\figurenum{4}
\centerline{
\includegraphics[width=8cm]{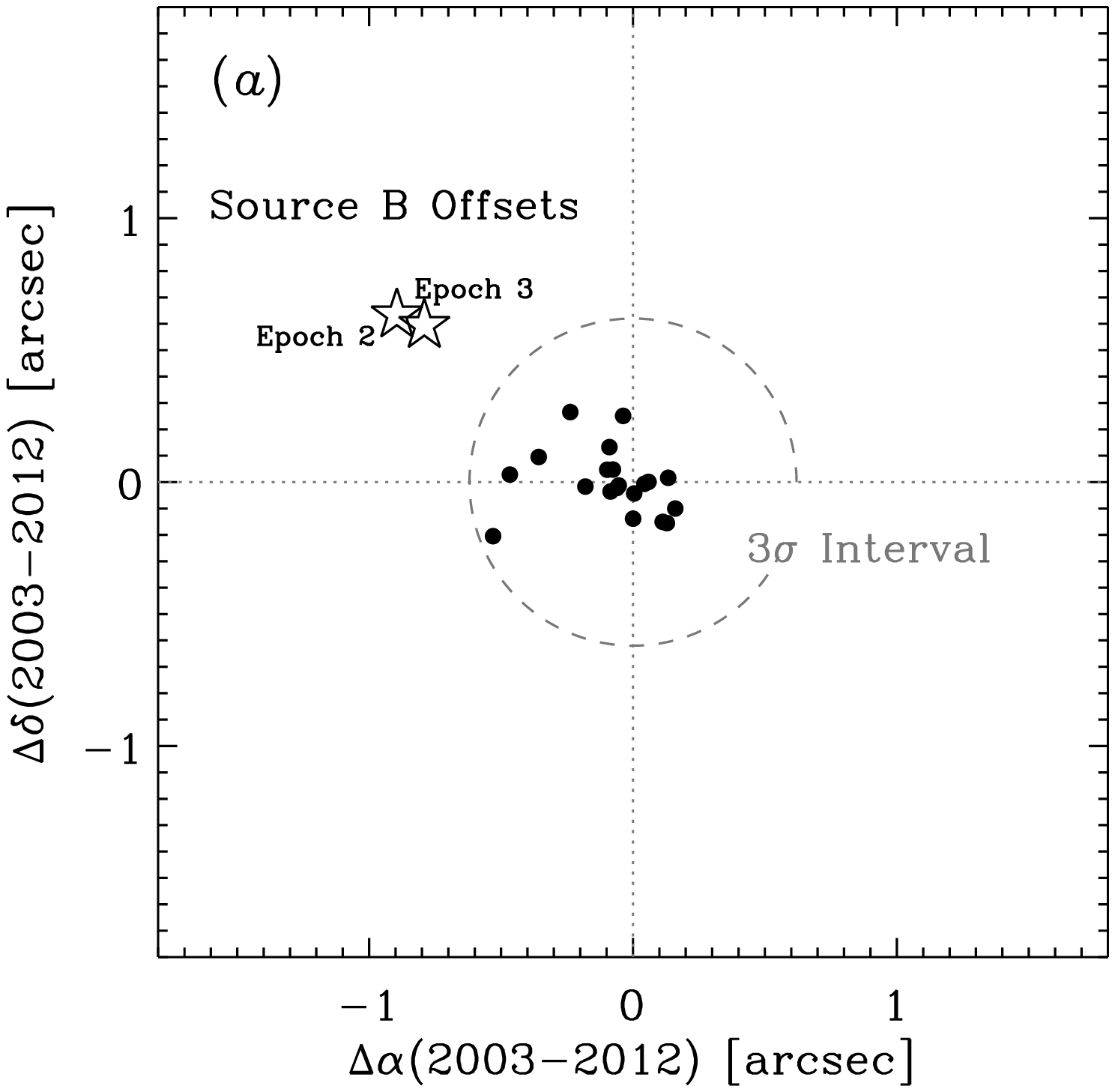}
\hfill
\includegraphics[width=8cm]{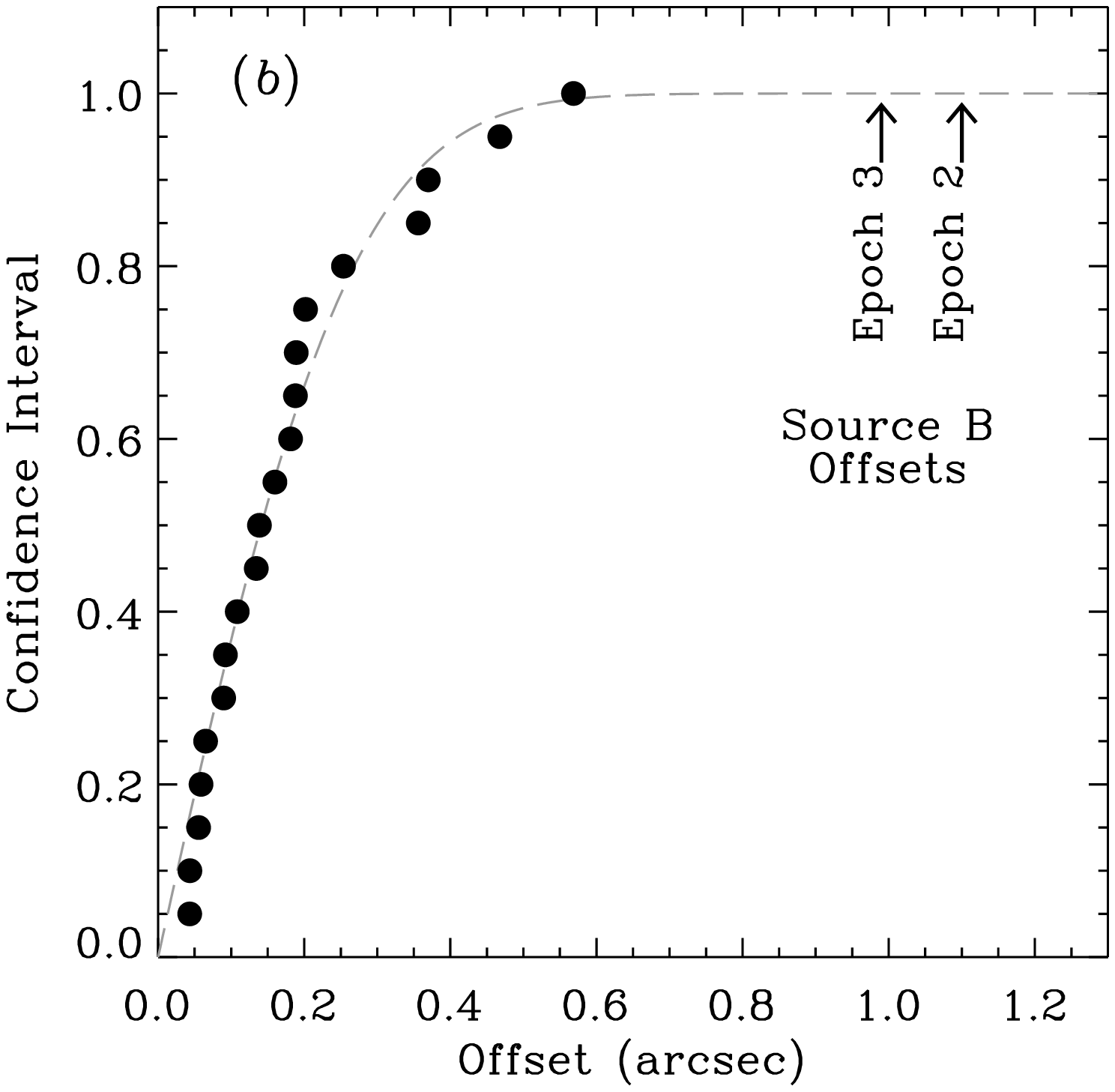}
}
\vspace{0.1in}
\caption{
($a$) \chandra\ positional offsets in declination $\Delta \delta$ and
right ascension $\Delta \alpha$ for 20 NGC~253 point-sources detected
in 2003 and our three 2012 observations ({\it filled circles\/}).  The dashed
circle shows the approximate 3$\sigma$ (99.7\%) confidence interval.  The offsets
between N-2003 and source~B in epochs 2 and 3 are indicated with open
stars.  
($b$) The cumulative offset distribution for the 20 matched point sources ({\it
filled circles\/}) and best-fitting error function ({\it dashed curve\/}).  The
cumulative offset and best-fitting error functions have been normalized to the
maximum error function, which provides direct estimates of confidence intervals.  The
offsets between N-2003 and source~B in epochs~2 and 3 are indicated.  Our
statistical analysis shows that N-2003 is unrelated to source~B at the
$\gg$99.99\% confidence level.
}
\end{figure*}

Assuming the {\it NuSTAR} variability is primarily due to variations in source
B, we can estimate its spectra by differencing the total nuclear region spectra
between bright and faint epochs.  Since the variations in the total {\it
NuSTAR} spectra appear to be consistent with only changes in spectral
normalization and not shape (see above), we assume that $S_i$, the total
nuclear region spectrum in epoch $i$, is described by
\begin{equation}
S_i \equiv S_{\rm const} + c_i S_B,
\end{equation}
where the constant $S_{\rm const}$ represents all the non-variable emission in
the nuclear region (e.g., due to sources A and C, as well as diffuse emission),
the constant $S_B$ represents the spectrum of source B, and $c_i$ represents
the variable normalization of source B at epoch $i$.  Indeed, Figure$3b$ shows
the difference spectra, which are well fit by an absorbed broken power-law
model with best-fit $N_{\rm H} = (1.6 \pm 0.5) \times 10^{23}$~cm$^{-2}$,
low-energy slope of $\Gamma_1  = 2.4 \pm 0.5$, break energy $E_{\rm break} =
7.9 \pm 0.9$~keV, and high-energy slope $\Gamma_2 = 3.9 \pm 0.4$ for both
cases, with only the normalization varying.  Fitting the difference spectra of
epoch~2 minus epoch~1 and epoch~3 minus epoch~1 separately yields similar
best-fit parameters and no improvement in spectral fit, implying the spectral
shape of source~B did not vary significantly between epochs 2 and 3.
Equation~1 then implies
\begin{equation}
c_iS_B = \frac{S_i - S_1}{1 - c_1/c_i}.
\end{equation}
From the {\it Chandra} observations, we know that $c_1/c_2 \approx 0.19$ and
$c_1/c_3 \approx 0.38$.  We note that no Fe line was necessary in our fits to
source~B implying that variable Fe emission is not seen in this source.  This
result alone likely rules out the possibility that source~B is a more luminous
version of the variable Fe reflection nebulae seen in the Galactic center,
since we would expect corresponding variability in the Fe emission line (e.g.,
Ponti \etal\ 2010).  

Integration of our spectral model and Eqn.~2 implies the 2--10~keV fluxes of
source~B were $\approx$7.4 and $3.0 \times 10^{-13}$~\flux, for observations 2
and 3, respectively.  These fluxes correspond to observed 2--10~keV
luminosities of $L_{\rm 2-10~keV} \approx 1.4 \times 10^{39}$ and $5.6 \times
10^{38}$~\lum, respectively.  The unabsorbed, intrinsic 2--10~keV luminosities
are therefore $L_{\rm 2-10~keV} \approx 5.1$ and $2.6 \times 10^{39}$~\lum,
respectively.  We note that the central black hole of NGC~253 has been
estimated to be $\approx$$5 \times 10^6$~\msol\ (Rodr{\'{\i}}guez-Rico \etal\
2006).  Such a black hole would have an Eddington luminosity of $L_{\rm
2-10~keV}^{\rm Edd} \approx 3 \times 10^{43}$~\lum.  For this approximation, we
assumed a 2--10~keV bolometric correction of $\approx$22.4, which corresponds
to the median bolometric correction for local AGN with $L_{\rm 2-10~keV}
\approx 10^{41}$--$10^{46}$~\lum\ (Vasudevan \& Fabian 2007).  If source~B were
powered by the black hole, then it would be accreting at $\sim$10$^{-4}$
Eddington.  AGN with these levels of $L_{\rm 2-10~keV}/L_{\rm Edd}$ typically
have spectral slopes of $\Gamma = 1.4 \pm 0.4$ (e.g., Shemmer \etal\ 2006;
Younes \etal\ 2011), much shallower than the spectrum measured for source~B.
This suggests that source~B is unlikely to be powered by a low-luminosity AGN.
Instead, the measured luminosities, lack of variability on the
$\approx$5--48~hr timescales, and \xray\ spectral shape of source~B closely
resemble the properties of binaries in the ultraluminous state, which include
ULXs that are likely to be stellar-mass black holes accreting above the
Eddington limit (e.g., Roberts~2007; Heil \etal\ 2009; Gladstone \etal\ 2009).
It is therefore likely that source~B is a ULX and not an AGN.  Given that
the number of ULXs in galaxies has been observed to correlate with galaxy-wide
SFR (e.g., Mineo \etal\ 2012), it would not be surprising to find a ULX
associated with the nuclear starburst of NGC~253.  Indeed, multiple bright
point sources and transient ULXs have also been observed in the nuclear
starburst in M82 (e.g., Feng \& Kaaret~2007).  Such a feature may be ubiquitous
among starburst galaxies.  At distances $\simgt$5--10~Mpc, even \chandra\ may
have difficulty distinguishing between such sources.

\subsection{Comparisons with the 2003 \chandra\ Observation}

As noted in $\S$~2 (see also Fig.~2$b$), the \chandra\ position of source~B is
$\sim$1~arcsec from the radio sources TH2 and TH4 (from Ulvestad \&
Antonucci~1997) and the galactic dynamical center (within the 1.2~arcsec
3$\sigma$ uncertainty radius; M{\"u}ller-S{\'a}nchez \etal\ 2010).  A similar
hard source located $\approx$0.4~arcsec from the dynamical center was
previously reported by M{\"u}ller-S{\'a}nchez \etal\ (2010), based on the
\chandra\ observation from 2003.  In Figure~2$e$, we show the nuclear region
image of the first 20~ks of the 2003 observation (to be equivalent to the depth
of our 2012 epochs), with the average locations of the 2012 sources circled.
It appears that the position of the 2003 hard near-nuclear source (labeled
``N-2003'' in Fig.~2$e$), located at ($\alpha$,$\delta$)$_{\rm J2000}$ =
00$^{\rm h}$~47$^{\rm m}$~33.12$^{\rm s}$, $-25^\circ$~17\arcmin~17.87\arcsec,
is offset from source~B by $\sim$1~arcsec ($\sim$19~pc) in the direction of
TH2/TH4, and the dynamical center.  We note that three other archival
\chandra\ exposures of NGC~253 (ObsID: 790, 969, and 383) were inspected for
evidence of N-2003 or source~B.  Although not formally detected, N-2003 is
visually apparent in a \hbox{4--7~keV} image from the shallower $\approx$14~ks
archival observation (ObsID: 969) that was conducted in December of 1999.  The
moderately-deep 45~ks observation (ObsID:~790), also conducted in December of
1999, had an aimpoint displaced $\approx$5~arcmin from the nuclear region,
which effectively blended the PSFs making it impossible to spatially measure
whether N-2003 or source~B were present.  Finally, the $\approx$2~ks exposure
conducted in August of 2000 (ObsID:~383) was too shallow to detect either
N-2003 or source~B.  We therefore choose to restrict further comparisons of our
2012 observations to the 2003 exposure.

Given the brightness of source~B and N-2003 ($\approx$50--300 net counts in the
2--7~keV band) and small off-axis angles with respect to the \chandra\ aim
points ($\simlt$0.5--1~arcmin), we expect the 90\% confidence positional errors
related to PSF centroiding to be $\approx$0.1--0.2~arcsec (based on Eqn.~13 of
Kim \etal\ 2007).  These positional errors are much smaller than the observed
offsets and strongly indicates that the two sources are unrelated.  Although
the \chandra\ astrometric frames were aligned to the 2003 observation (see
$\S$2 for details), some non-negligible error related to image registration is
expected.  To test further whether the offsets between N-2003 and source~B are
statistically significant, given our best image registrations, we matched
\chandra\ sources detected in the 2003 observation to counterparts detected in
each of the three 2012 observations, excluding source~B and N-2003.  For our
matching, we required that sources in each catalog have $\simgt$40 net counts
in the 2--7~keV band and be located within $\approx$3~arcmin of the nucleus, so
that the signal-to-noise ratios and \chandra\ PSFs are comparable to those of
N-2003 and source~B.  Two sources were considered to match if they were
separated by $<$2~arcsec.  In Figure~4$a$, we show the right ascension
($\alpha$) and declination ($\delta$) offsets for all 20 matches and highlight
the offsets between the N-2003 and source~B in epochs 2 and 3 (source~B was not
detected in epoch~1).  We find that the offsets between N-2003 and source~B are
1.10 and 0.99~arcsec for the epoch~2 and 3 positions, respectively; both are much
larger than the maximum offset of the 20 matches (0.6~arcsec).  In Figure~4$b$,
we show the cumulative number of matches as a function of offset with a
best-fit error function, which describes well the cumulative offset
distribution.  The data in Figure~4$b$ have been normalized by the error
function maximum to allow for direct estimates of the confidence intervals.
Our error function suggests that the source~B offsets are both significant at
the $\simgt$99.99\% confidence levels.  Taken together, we infer that source~B
is unrelated to N-2003.

%
%
\begin{figure}
\figurenum{5}
\centerline{
\includegraphics[width=8.9cm]{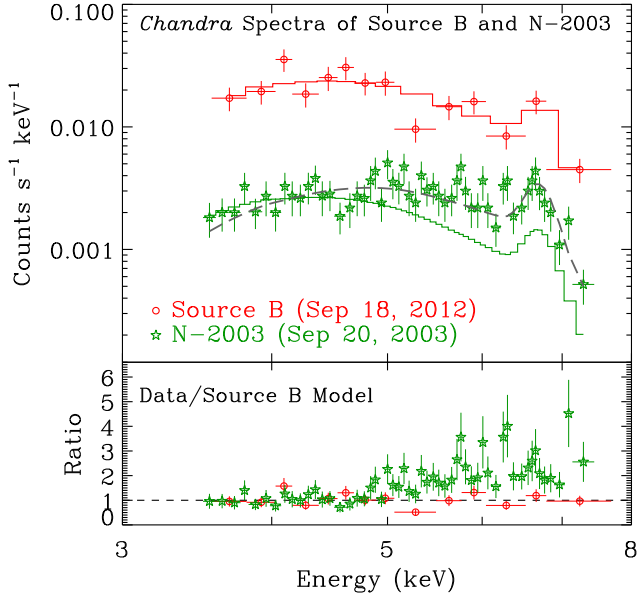}
}
\vspace{0.1in}
\caption{
\chandra\ spectra of source~B in epoch~2 ({\it red circles\/}) and N-2003 ({\it
green stars\/}).  For visualization purposes, the normalization of source~B has
been scaled upwards by a factor of four.  The solid red curve shows the
best-fit spectral model for source B from Figure~3$b$, with an Fe component
added.  The solid green curve shows the same model renormalized to the
$\simlt$5~keV spectrum of N-2003.  The bottom panel displays the ratio between
the data and the 2012 model (solid curves in the top panel), illustrating that
N-2003 has a harder spectrum than source~B.  The \chandra\ spectrum of N-2003
can be fit well by an absorbed power-law with $N_{\rm H} \approx 2.8 \times
10^{23}$~cm$^{-2}$ and $\Gamma \approx 1.9$ ({\it dashed curve} in the top
panel), which differs from the $\Gamma \approx 3.1$ slope of source~B.  
}
\end{figure}

The above analysis shows convincingly that N-2003 is unrelated to source~B
implying that at least one of these sources is not associated with the central
black hole.  Our spectral constraints, presented in $\S$3.2, show that source~B
is most likely to be a ULX.  On the other hand N-2003 is offset from the
dynamical center by only $\approx$0.4~arcsec (i.e., within the 1.2~arcsec
3$\sigma$ uncertainty of the dynamical center), indicating that it is a better
candidate for a ``true'' nuclear \xray\ point source that may be an AGN (see
Fig.~2$e$).  In Figure~5, we show the 3--8~keV \chandra\ spectra of N-2003 and
source~B (when at its peak in epoch~2).  We note that the morphology of the
diffuse 3--8~keV \chandra\ emission in the nuclear region is complex and
appears brightest in the immediate vicinity of N-2003 and source~B (see
Fig.~2$b$--2$e$).  As such, detailed modeling of the background spectrum
associated with the diffuse emission in this region is difficult.  To mitigate
this limitation, we restricted our \chandra\ spectral analysis to energies
above 3~keV to exclude strong continuum contributions from the hot ISM and
added an Fe component at 6.7~keV to account for the gas emission line (see
$\S$3.1 for motivation).  Given that the \nustar\ difference spectra of
source~B do not show a strong Fe line (see Fig.~3$b$), we do not expect that
this component is intrinsic to source~B; however, we are less certain about the
nature of the Fe line associated with N-2003.  The best-fitting spectral model
for source~B, presented in $\S$3.1, renormalized to the 3--5~keV flux of
N-2003, is obviously far too steep to fit the higher-energy 5--8~keV spectrum
of N-2003, supporting the conclusion that source~B and N-2003 are different,
despite having similar 2--10~keV fluxes.  As a caveat, we note that \xray\
binaries accreting at near-Eddington rates can have variable spectral
properties without substantial changes in luminosity (e.g., GRS~1915+105;
Vierdayanti \etal\ 2010), implying that the harder spectrum of N-2003 does not
rule out a ULX origin similar in nature to source~B.  The
\chandra\ spectrum of N-2003 can be fit well using an absorbed power-law model
with $N_{\rm H} = (2.8 \pm 0.6) \times 10^{23}$~cm$^{-2}$ and $\Gamma = 1.9 \pm
0.6$, and an Fe line fixed at 6.7~keV.  If we used only the \chandra\ data to
fit the spectrum of source~B, we find $N_{\rm H} = (1.9 \pm 0.8) \times
10^{23}$~cm$^{-2}$ and $\Gamma = 3.1 \pm 1.0$, consistent with the values found
from our \nustar\ analysis in $\S$3.2.  Although the fitting parameters of
N-2003 and source~B are formally consistent, the spectrum of N-2003 is harder
than that of source~B (see bottom panel of Fig.~5), and more consistent with an
AGN ($\Gamma \approx$~1.5--2.2; see further discussion above).  

Due to its close proximity to the dynamical center and its harder spectrum, it
is plausible that N-2003 is an AGN that was in a low state during our 2012
monitoring campaign.  However, we cannot currently rule out a ULX nature for
N-2003.  If N-2003 were an AGN, the $\Gamma \approx 1.9$ power-law spectrum
would extend well into the \nustar\ bandpass out to beyond $\approx$40~keV.
The observed \hbox{2--7~keV} flux of N-2003 is 1.1 times that of source~B in
epoch~2.  However, if N-2003 were an AGN, then we predict that it would have
had a \hbox{10--20~keV} flux $\simgt$2.5 times higher than that of source~B at
its peak in epoch~2.  Therefore, future monitoring of NGC~253 with both
\chandra\ and \nustar\ would be able to resolve the degeneracy between the AGN
and ULX nature of N-2003 if caught in a high state.

%
\section{Summary}
%

We performed nearly simultaneous \nustar\ and \chandra\ monitoring of the
nearby starburst galaxy NGC~253 over three observational epochs: beginning 2012
September 1, 2012 September 15, and 2012 November 16.  We find that the
7--20~keV nuclear region flux was elevated over the epoch~1 level by factors of
$\approx$1.7 and $\approx$1.4 in epochs~2 and 3, respectively.  Our \chandra\
observations show that a single variable source, which we call source~B (see
Fig.~2$b$--2$e$), was responsible for driving this variation.  The \nustar\
difference spectra (i.e., subtracting epoch~1 from the bright states in
epochs~2 and 3) allows us to study the spectrum of source~B over the 3--20~keV
energy range.  Source~B has a peak observed 2--10~keV luminosity of
$\approx$$1.4 \times 10^{39}$~\lum\ (estimated unabsorbed, intrinsic $L_{\rm
2-10~keV} \approx 5.1 \times 10^{39}$~\lum) and is fit well by a broken
power-law model with spectral slopes and a break energy within the range of
values characteristic of ULXs and not AGN.  

A previous \chandra\ observation in 2003 revealed a hard \xray\ source
``N-2003'' that had a similar 2--7~keV flux to that observed for source~B in
epoch~2; however, the position of N-2003 is displaced from source~B by
$\approx$1~arcsec.  The high-precision imaging of \chandra\ allows us to show
at the $\gg$99.99\% confidence level that N-2003 is unrelated to source~B.
Further examination of the position of N-2003 indicates that it is a better AGN
candidate than source~B; however, the \chandra\ spectrum alone cannot rule out
that N-2003 may be a second ULX.  We predict that if N-2003 were an AGN, then a
\nustar\ observation would have yielded a 10--20~keV flux that was $\simgt$2.5
times higher than that of source~B at its peak brightness.  In light of this,
future monitoring with \chandra\ and \nustar\ would be able to break the
degeneracy between a ULX and AGN nature of N-2003, if it returns to a bright
state.

\acknowledgements

We thank the anonymous referee for their helpful comments, which have
improved the quality of this paper.  This work was supported under NASA
Contract No. NNG08FD60C, and  made use of data from the {\it NuSTAR} mission, a
project led by  the California Institute of Technology, managed by the Jet
Propulsion  Laboratory, and funded by the National Aeronautics and Space
Administration. We thank the {\it NuSTAR} Operations, Software and  Calibration
teams for support with the execution and analysis of  these observations.  This
research has made use of the {\it NuSTAR}  Data Analysis Software (NuSTARDAS)
jointly developed by the ASI Science Data Center (ASDC, Italy) and the
California Institute of  Technology (USA).  We thank the \chandra\ \xray\
Center staff for providing faster than usual processing of the \chandra\ data.

%

%

\end{document}